\documentclass[a4paper,11pt]{article}
\pdfoutput=1 % if your are submitting a pdflatex (i.e. if you have
\usepackage{jinstpub} % for details on the use of the package, please
\title{\boldmath Technical instrumentation R\&D for
ILD SiW ECAL large scale device}

\author[a,1]{V.~Balagura\note{Corresponding author.}}

\affiliation[a]{Laboratoire Leprince-Ringuet (LLR) - \'Ecole polytechnique, avenue Fresnel, Palaiseau, France}

\emailAdd{balagura@llr.in2p3.fr}

\abstract{Calorimeters with silicon detectors have many unique features and
  are proposed for several world-leading experiments. We describe the R\&D
  program of the large scale detector element with up to 12 000 readout
  channels for the International Large Detector (ILD) at the future $e^+e^-$
  ILC collider. The program is focused on the readout front-end electronics
  embedded inside the calorimeter. The first part with 2 000 channels and two
  small silicon sensors has already been constructed, the full prototype is
  planned for the beginning of 2018.}


\collaboration[c]{on behalf of SiW ECAL ILD collaboration \qquad \qquad
  \includegraphics[height=12mm]{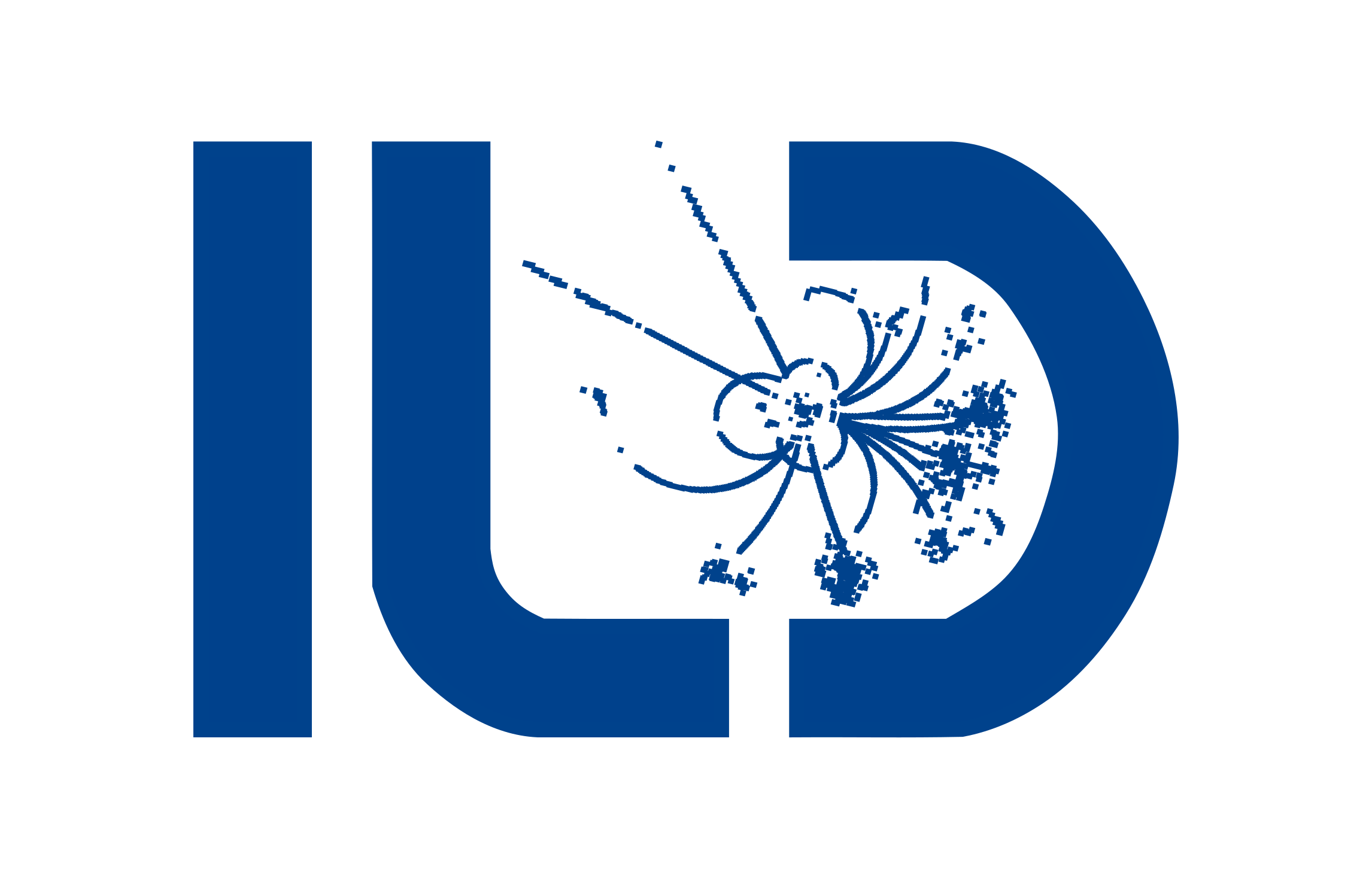}}

\proceeding{Calorimetry for the High Energy Frontier, CHEF2017\\
  2 - 6 October 2017\\
  Lyon, France}

\begin{document}
%\notoc
\maketitle
\flushbottom

\section{SiW ECAL in the ILD}
\label{sec:intro}

The calorimetry system of the detector at the future $e^+e^-$ International
Linear Collider (ILC) is optimized for the precision measurement of quarks,
gluons and tau's 4-momenta using a Particle Flow Algorithm
(PFA)~\cite{pfa1,pfa2}. The most demanding requirements are for the
electromagnetic calorimeter (ECAL) which should both well separate individual
electromagnetic showers and distinguish them from hadrons.

In terms of PFA requirements the most promising is the silicon-tungsten (SiW)
ECAL technology with the pixel size of about 5$\times$5~mm$^2$. In this paper
we shall discuss the SiW ECAL for the International Large
Detector (ILD)~\cite{ild} at the ILC. Depending on the ILD size and the number of ECAL
layers, the total number of ECAL channels can be between 60 -- 100
million. In spite of having so many channels, the complexity of the
commissioning and the operation will be significantly reduced compared to
other technologies, thanks to the perfect intrinsic linearity of the silicon
PIN diodes, the time stability, the uniformity of the gains across channels
and the ease of calibration. Overall, this should allow to achieve the lowest
ECAL systematics.

\begin{figure}[htbp]
\centering
\includegraphics[width=6.8cm]{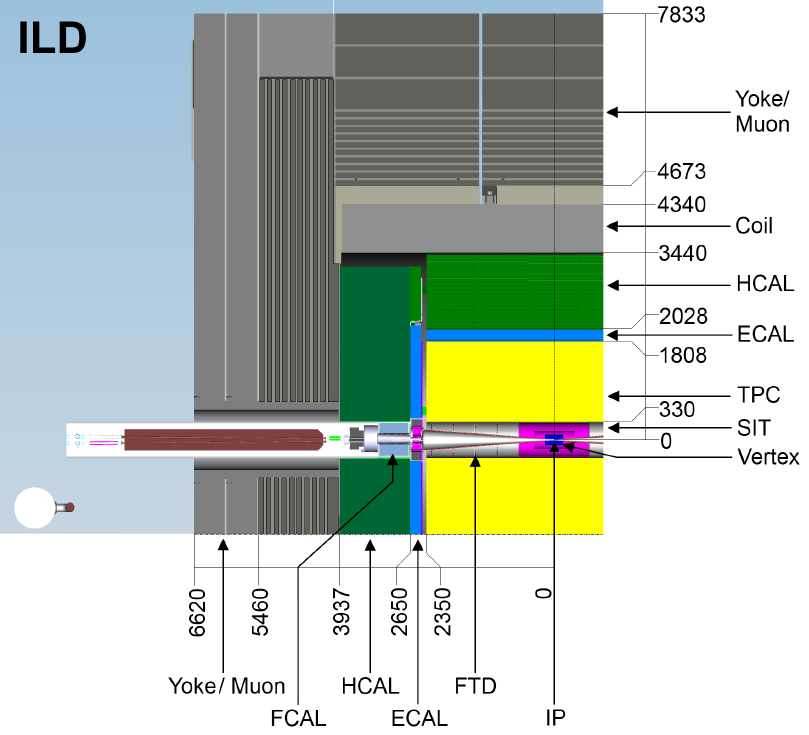}\hspace{0.6cm}
\includegraphics[width=6.6cm,trim=0 -1cm 0 0]{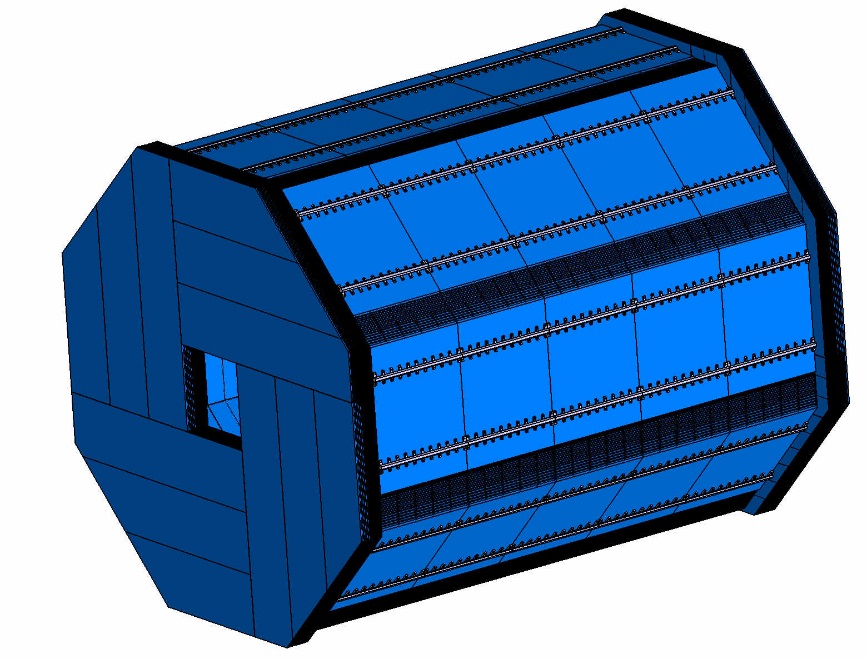}
\caption{\label{ild} Left: quadrant view of the ILD detector, taken
  from~\cite{ild}.  The interaction point is in the lower right corner,
  dimensions are in mm. ECAL is shown in blue. Right: octagonal ECAL barrel
  and two endcaps.}
\end{figure}

\begin{figure}[htbp]
\centering
\includegraphics[width=6.8cm]{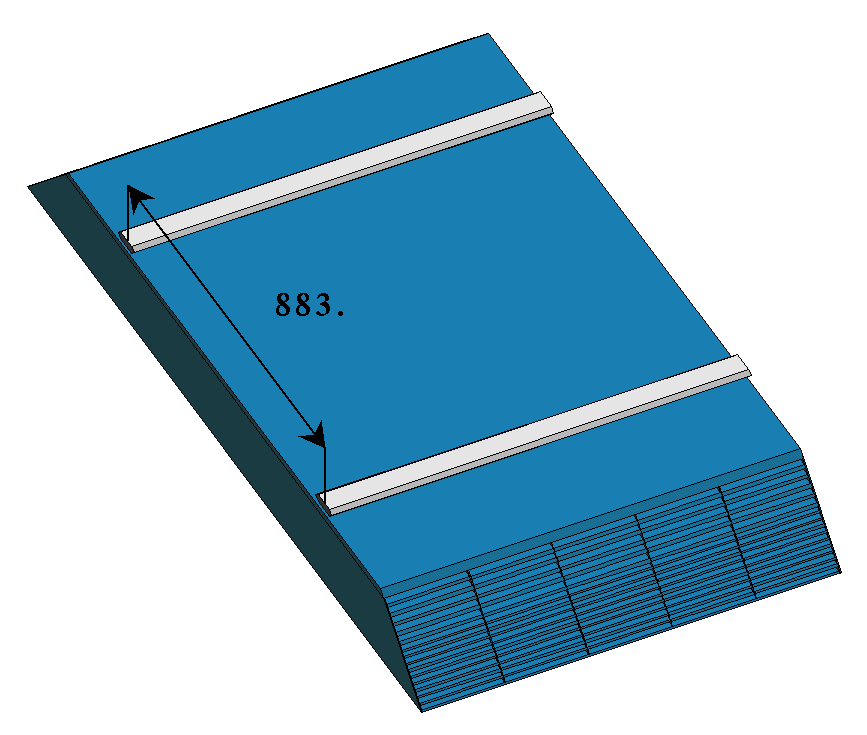}\hspace{0.4cm}
\includegraphics[width=6.8cm,trim=0 -2cm 0 0,clip]{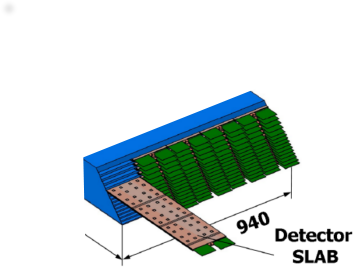}
\caption{\label{module} Left: one empty barrel module without active detector
  elements (slabs). Right: insertion of the slabs into alveoli (only the end
  of the module is shown).}
\end{figure}

The ILD ECAL is placed between the hadron calorimeter (HCAL, see Fig.~\ref{ild},
left) and the time projection chamber (TPC) with the surrounding silicon
external tracker (SET, not shown).
ECAL has an octagonal barrel part and two endcaps, as shown in Fig.~\ref{ild},
right. The former has 8 independent sides not touching each other.  Every side
is divided longitudinally along the beam axis into 5 modules attached to HCAL
on common rails.  One barrel module without active detector elements is shown
in Fig~\ref{module}, left.
It is made of a carbon-fiber structure holding every second layer of the
tungsten absorber.  It has cavities called alveoli into which the active
detectors called slabs slide in, as shown in Fig.~\ref{module}, right.
\begin{figure}[htbp]
  \centering
\includegraphics[width=14cm]{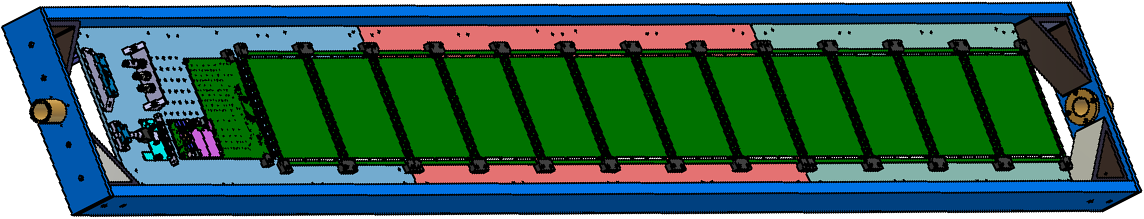}
\caption{\label{long_slab} Long slab prototype design: 12 PCBs with glued
  small silicon sensors are readout from the left side. External cables pass
  through the hole positioned in the middle of the left side on the rotation
  axis. Three colors of the base plate distinguish its different parts,
  installation of one, two or all three parts allows to vary the prototype
  length. The base plate is made of polyvinyl chloride (PVC), the side frames,
  the edges, the mounting supports in the corners and the base plate supports
  on the opposite side are made of aluminum, the rotating part is built from
  stainless steel.}
\end{figure}
The slabs carry in the middle the other layers of tungsten wrapped in
carbon-fiber. Both sides of the absorber are equipped with the silicon
matrices of PIN diode pixels and the front-end electronics. The silicon
sensors are glued pixel-by-pixel to the readout printed circuit boards (PCBs)
with a conductive epoxy.  PCBs are connected in-line and read out from one
slab end. All PCBs except the last one have a square shape.  The endcaps are
divided into 4 quadrants, each having 3 or 2 modules depending on the ILD
size. The endcap modules are similar to the barrel ones but have a different
shape, in particular, the slabs there are longer. In addition, the most
forward ECAL endcap regions will be instrumented with the so-called `endcap
rings'' (not shown in Fig.~\ref{ild}). Their design is not yet finalized.

Currently, in addition to the ECAL with 30 layers described in the ILD TDR,
there are design options with 22 layers, with 23\% smaller radius and with
thicker (725~$\mu$m instead of 525~$\mu$m) and larger silicon sensors made
from 8 instead of 6 inch wafers.

% Videau structure, odd N modules, no cracks in endcaps

\section{CALICE physical and technological SiW ECAL prototypes}

In 2004--2011 the first so-called ``physical'' prototype of SiW ECAL with 30
layers has been built and successfully tested within the CALICE
collaboration~\cite{phys}. It was designed to prove the physical PFA
principles in beam tests. The electronics was not embedded inside the
calorimeter but placed alongside.

After successful completion of the physical prototype program, the emphasis
has moved to the technological realization of the detector scalable to the ILD
ECAL. This second generation prototype is called ``technological''. The
electronics is embedded in the active layer.  It is based on a dedicated 64
channel front-end chip SKIROC~\cite{skiroc}.
%% Each channel has 15 memory slots called SCA digitized by
%% 12 bit Analog-to-Digital-Converter (ADC) linear in the range of about 1--1500
%% MIPs.
Since in the ILD there is no central trigger, every SKIROC chip has
auto-trigger capabilities, and the full event is built using bunch crossing
(BX) time stamps.

In the current design, the ILC collides bunch trains during 1~msec and then stays
idle for 199~msec. The latter period is used to readout the data stored in
SKIROC. To save the power and to simplify the cooling, the front-end
electronics are switched off after each readout cycle. This is called "power
pulsing". In such a mode one SKIROC channel consumes only about 27~$\mu W$. In
practice, however, the electronics should be switched on about 1~msec
before the bunch trains so that the transition processes (causing
e.g. pedestal drifts) finish well in advance.

One PCB of the current prototype has 16 SKIROC chips and 1024 channels. It
serves 4 silicon sensors each segmented into 256 pixels of
5.5$\times$5.5~mm$^2$ size. Up to now, seven such single PCB detectors have
been built. They have been tested with cosmics, charge injection and in 3 beam
test campaigns: at CERN in November 2015 and June 2016 and in DESY in June
2017. Every PCB was readout independently. In the test beams they formed a
mini-ECAL with 7 layers. Overall, they have demonstrated a good
performance~\cite{instr,chef} though there is still some room for
improvement of the readout electronics.

Single PCB detectors with size of about $18\times18$~cm$^2$ are called
``short'' slabs to distinguish them from multi-PCB ``long'' slabs required in
the ILD. Their length, depending on the ILD size, can be as large as 1.5~m
(2~m) in the barrel (endcap).  This paper describes the R\&D program aiming to
build the first ``long'' slab prototype.  Based on the learned experience from
the ``short'' slabs, the current R\&D is focused only on the performance of
the readout electronics which is considered as the most difficult task. In
particular, the future ``long'' slab prototype should verify:
\begin{itemize}
\item clock and signal propagation along the ``long'' slab and their distortions,
\item low and high voltage distribution,
\item ability to read large data volumes and the integrity of data,
\item PCB electrical interconnections.
\end{itemize}

For simplicity, the prototype will leave out e.g. the ILD constraints on the
sizes and the mechanical tolerances.  Contrary to the ILD slab, the prototype
will have no absorber and will have only one layer of PCB (the second should
be identical).

The scheme of the long slab prototype together with the holding mechanical
structure is shown in Fig.~\ref{long_slab}. Up to 12 PCBs will be connected
in-line. Similarly to the existing ``short'' slabs, the long slab will be
communicated from one end via the so-called adapter and DIF boards, the latter
being connected to the computer via a concentrator board GDCC (not shown). It
will be possible to configure individually every chip of the prototype,
for example, to perform individual masking of the channels.

In the ``short'' slabs, the interconnection between the PCB and the adapter
board is performed by permanent soldering of a dedicated flexible flat kapton
cable. The same method of interconnections is currently foreseen in the ILD
for reliability. Here, due to a magnetic field there will be a Lorentz force
acting on wires which feed currents to the front-end electronics. Due to the
power pulsing, the currents will switch on and off at a frequency of
5~Hz. This will create vibrations, in particular, in the interconnections.
The long slab prototype will not address the question of 100\% reliability of
the interconnections required for the ILD.  For simplicity, the
interconnections will be performed with Flat Flexible Cable (FFC)
connectors. This will give the flexibility to freely connect and disconnect
the PCBs which might be important for the future tests.

Each PCB will be equipped with one small silicon sensor having 4$\times4$
pixels. This should be sufficient to study the physical performance of the
detector. All 1024 channels of PCB will be checked with the charge injection.

The PCBs will be added to the long slab one-by-one, each time verifying the
performance.  In case of problems, the maximal allowed length of the current
long slab prototype will be determined. Otherwise, the long slab will have 12
PCBs, as required for the endcap in the ILD with the maximal size.  The base
frame will contain 3 pieces, shown in Fig.~\ref{long_slab} by different
colors.  By installing only one, two or all three pieces, one can vary the
slab length, such as it can hold 2, 8 or 12 PCBs.  The performance of the slab
will be tested with $^{90}_{38}$Sr and $^{137}_{55}$Cs radioactive sources,
cosmics, charge injection and in a beam test planned in 2018.

% interconncetions: ILD soldering vs connector

\section{Current status of the ILD long slab prototype}

The design of the long slab is finished. It can be rotated around its middle
axis and fixed at any position. Vertical and horizontal orientations are
foreseen for the beam and cosmic tests, respectively. The prototype length is
3.22~m, which is longer than the ILD slab because of the margins and an extra
space needed for the current PCB interconnections. With the weight slightly
less than 30~kg, the prototype can be handled by two people.  The base frame
is rigid. When held at two ends, the mechanical bending is less than 1~mm
within one PCB area. The silicon sensor should be perfectly flat, therefore
the PCB is fixed to the frame using a flexible support printed at a 3D
printer. This prevents the bending to be propagated to the PCB and the
sensor. % The PCB fixation is done only at its technical edges.

Access to the electronic boards is foreseen from both sides, the base frame
has special holes under each PCB for that. The boards are electrically
isolated to reduce noise. The prototype will be light-tight since the silicon
detectors are sensitive to ambient light.

Even a small resistivity of low voltage (LV) power lines results in voltage
drops due to large currents. They might become sizable at the far end of the
slab (several hundred mV) if the power is distributed through the PCBs along
the slab. Therefore, for simplicity, the PCBs will be powered not sequentially
but by individual cables of the same length. There will be two such cables for
LV and a ground, and their length will be determined by the last PCB at the
slab end.

The high voltage (HV) will be supplied to the silicon sensors by a separate
cable glued at one end to a HV pad of the sensor. The other end will be
possible to disconnect from a HV source, if it will be necessary to change the
PCB with the sensor.

\begin{figure}[htbp]
  \centering
\includegraphics[width=9cm]{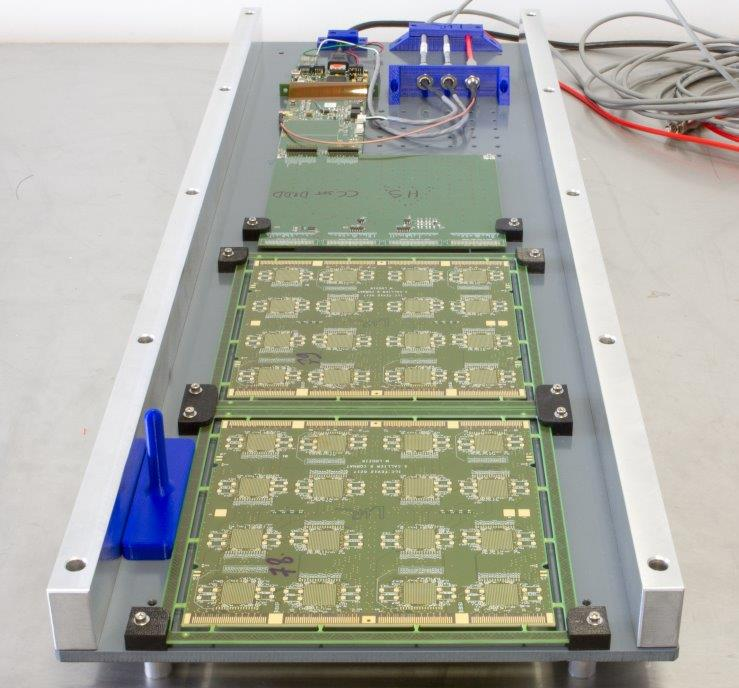}
\caption{\label{photo} Two first PCBs installed in the long slab prototype. At
  the moment when this photo was taken, the PCBs were not equipped with the
  electronics and the sensors. At present, they are already operational and
  ready for the first tests.  The adapter and DIF readout boards are visible
  at the top of the photo. The grey PCB supports installed at their corners are
  flexible, so that PCB and the silicon sensors remain flat even if the base
  frame bends.  This is not critical for the installed small $4\times4$ pixel
  detectors but foreseen for a future tests with large sensors. The left blue
  wedge visible on the left between the PCB and the side frame is used for the
  alignment.}
\end{figure}

The first two PCBs with the glued small sensors have been already installed in
the long slab prototype (see Fig.~\ref{photo}) and prepared for the first
tests. All other PCBs and sensors are available. They will be installed
one-by-one and tested. The full long slab prototype is due in the beginning of
March 2018.

\section{Conclusions}

In this paper we have described the plans and the first steps to build the
first prototype of the ILD SiW ECAL detector element with up to 12 288
channels. It will contain up to 12 PCBs connected in-line and forming the
so-called ``long'' slab.  The main focus is on the electronics readout which
is considered as the most difficult part of the R\&D. The prototype will be
controlled and read out from one end using existing adapter, DIF and GDCC
boards. Every PCB has 16 SKIROC chips and 1024 channels. It will be
equipped with the small silicon sensor with $4\times4$ pixels. This should be
sufficient to check the physical performance of the prototype. The other
channels will be checked with the charge injection available in SKIROC.

The first part of the prototype with 2048 channels have been already
built. The full prototype is due in the beginning of 2018.

\acknowledgments

Supported by the H2020 project AIDA-2020, GA no. 654168.


\end{document}